\documentclass{article}
\usepackage[utf8]{inputenc}
\usepackage{graphicx}
\begin{document}

%
%
%
%CORRECT LASZLO with diacritics

\title{A Lost Croatian Cybernetic Machine Translation Program 
}

\author{Sandro Skansi*       \and
        Leo Mr\v si\' c           \and 
        Ines Skelac 
}

\author{Sandro Skansi*  \\
	 University of Zagreb\\ University Department of Croatian Studies   \\
	 sskansi@hrstud.hr
	\and 
	Leo Mr\v si\' c \\
	Algebra University College \\
	leo.mrsic@algebra.hr
	\and
	Ines Skelac \\
	Algebra University College \\
	ines.skelac@racunarstvo.hr
	}

\maketitle

\begin{abstract}
\noindent We are exploring the historical significance of research in the field of machine translation conducted by Bulcsu Laszlo, Croatian linguist, who was a pioneer in machine translation in Yugoslavia during the 1950s. We are focused on two important seminal papers written by members of his research group from 1959 and 1962, as well as their legacy in establishing a Croatian machine translation program based around the Faculty of Humanities and Social Sciences of the University of Zagreb in the late 1950s and early 1960s. We are exploring their work in connection with the beginnings of machine translation in the USA and USSR, motivated by the Cold War and the intelligence needs of the period. We also present the approach to machine translation advocated by the Croatian group in Yugoslavia, which is different from the usual logical approaches of the period, and his advocacy of cybernetic methods, which would be adopted as a canon by the mainstream AI community only decades later.


\end{abstract}

\newpage 
\section{Beginnings of Machine Translation and Artificial Intelligence in the USA and USSR}
\label{intro}
In this paper, we are exploring the historical significance of Croatian machine translation research group. The group was active in 1950s, and it was conducted by Bulcsu Laszlo, Croatian linguist, who was a pioneer in machine translation during the 1950s in Yugoslavia.

To put the research of the Croatian group in the right context, we have to explore the origin of the idea of machine translation. The idea of machine translation is an old one, and its origin is commonly connected with the work of Rene Descartes, i.e. to his idea of universal language, as described in his letter to Mersenne from 20.xi.1629 \cite{Descartes1}. Descartes describes universal language as a simplified version of the language which will serve as an “interlanguage” for translation. That is, if we want to translate from English to Croatian, we will firstly translate from English to an “interlanguage”, and then from the “interlanguage” to Croatian. As described later in this paper, this idea had been implemented in the machine translation process, firstly in the Indonesian-to-Russian machine translation system created by Andreev, Kulagina and Melchuk from the early 1960s.

In modern times, the idea of machine translation was put forth by the philosopher and logician Yehoshua Bar-Hillel (most notably in \cite{Hillel2} and \cite{Hillel3}), whose papers were studied by the Croatian group. Perhaps the most important unrealized point of contact between machine translation and cybernetics happened in the winter of 1950/51. In that period, Bar-Hillel met Rudolf Carnap in Chicago, who introduced to him the (new) idea of cybernetics. Also, Carnap gave him the contact details of his former teaching assistant, Walter Pitts, who was at that moment with Norbert Wiener at MIT and who was supposed to introduce him to Wiener, but the meeting never took place \cite{Hutchkins4}. Nevertheless, Bar-Hillel was to stay at MIT where he, inspired by cybernetics, would go to organize the first machine translation conference in the world in 1952 \cite{Hutchkins4}.

The idea of machine translation was a tempting idea in the 1950s. The main military interest in machine translation as an intelligence gathering tool (translation of scientific papers, daily press, technical reports, and everything the intelligence services could get their hands on) was sparked by the Soviet advance in nuclear technology, and would later be compounded by the success of Vostok 1 (termed by the USA as a “strategic surprise”). In the nuclear age, being able to read and understand what the other side was working on was of crucial importance \cite{RussellNorvig5}. 
Machine translation was quickly absorbed in the program of the Dartmouth Summer Research Project on Artificial Intelligence in 1956 (where Artificial Intelligence as a field was born), as one of the five core fields of artificial intelligence (later to be known as natural language processing). One other field was included here, the “nerve nets” as they were known back then, today commonly known as artificial neural networks. What is also essential for our discussion is that the earliest programming language for artificial intelligence, Lisp, was invented in 1958 by John McCarthy \cite{Norvig6}.
But let us take a closer look at the history of machine translation. In the USA, the first major wave of government and military funding for machine translation came in 1954, and the period of abundancy lasted until 1964, when the National Research Council established the Automatic Language Processing Advisory Committee (ALPAC), which was to assess the results of the ten years of intense funding. The findings were very negative, and funding was almost gone \cite{RussellNorvig5}, hence the ALPAC report became the catalyst for the first “AI Winter”.

One of the first recorded attempts of producing a machine translation system in the USSR was in 1954 \cite{Piotr7}, and the attempt was applauded by the Communist party of the Soviet Union, by the USSR Committee for Science and Technology and the USSR Academy of Sciences. The source does not specify how this first system worked, but it does delineate that the major figures of machine translation of the time were N. Andreev of the Leningrad State University, O. Kulagina and I. Melchuk of the Steklov Mathematical Institute. There is information on an Indonesian-to-Russian machine translation system by Andreev, Kulagina and Melchuk from the early 1960s, but it is reported that the system was ultimately a failure, in the same way early USA systems were. The system had statistical elements set forth by Andreev, but the bulk was logical and knowledge-heavy processing put forth by Kulagina and Melchuk. The idea was to have a logical intermediate language, under the working name “Interlingua”, which was the connector of both natural languages, and was used to model common-sense human knowledge. For more details, see \cite{Piotr7}.

In the USSR, there were four major approaches to machine translation in the late 1950s \cite{Mulic8}. The first one was the research at the Institute for Precise Mechanics and Computational Technology of the USSR Academy of Sciences. Their approach was mostly experimental and not much different from today's empirical methods. They evaluated the majority of algorithms known at the time algorithms over meticulously prepared datasets, whose main strength was data cleaning, and by 1959 they have built a German-Russian machine translation prototype. The second approach, as noted by Muli\' c \cite{Mulic8}, was championed by the team at the Steklov Mathematical Institute of the USSR Academy of Sciences led by A. A. Reformatsky. Their approach was mainly logical, and they extended the theoretical ideas of Bar-Hillel \cite{Hillel3} to build three algorithms: French-Russian, English-Russian and Hungarian-Russian. The third and perhaps the most successful approach was the one by A. A. Lyapunov, O. S. Kulagina and R. L. Dobrushin. Their efforts resulted in the formation of the Mathematical Linguistics Seminar at the Faculty of Philology in Moscow in 1956 and in Leningrad in 1957. Their approach was mainly information-theoretic (but they also tried logic-based approaches \cite{Mulic8}), which was considered cybernetic at that time. This was the main role model for the Croatian efforts from 1957 onwards. The fourth, and perhaps most influential, was the approach at the Experimental Laboratory of the Leningrad University championed by N. D. Andreev \cite{Mulic8}. Here, the algorithms for Indonesian-Russian, Arabic-Russian, Hindu-Russian, Japanese-Russian, Burmese-Russian, Norwegian-Russian, English-Russian, Spanish-Russian and Turkish-Russian were being built. The main approach of Andreev's group was to use an intermediary language, which would capture the meanings \cite{Mulic8}. It was an approach similar to KL-ONE, which would be introduced in the West much later (in 1985) by Brachman and Schmolze \cite{Desc8}. It is also interesting to note that the Andreev group had a profound influence on the Czechoslovakian machine translation program \cite{Pogb10}, which unfortunately suffered a similar fate as the Yugoslav one due to the lack of funding.

Andreev's approach was in a sense "external". The modelling would be statistical, but its purpose would not be to mimic the stochasticity of the human thought process, but rather to produce a working machine translation system. Kulagina and Melchuk disagreed with this approach as they thought that more of what is presently called "philosophical logic" was needed to model the human thought process at the symbolic level, and according to them, the formalization of the human thought process was a prerequisite for developing a machine translation system (cf. \cite{Piotr7}). We could speculate that sub-symbolic processing would have been acceptable too, since that approach is also rooted in philosophical logic as a way of formalizing human cognitive functions and is also "internal" in the same sense symbolic approaches are.

There were many other centers for research in machine translation: Gorkovsky University (Omsk), 1st Moscow Institute for Foreign Languages, Computing Centre of the Armenian SSR and at the Institute for Automatics and Telemechanics of the Georgian SSR \cite{Mulic8}. It is worthwhile to note that both the USA and the USSR had access to state-of-the-art computers, and the political support for the production of such systems meant that computers were made available to researchers in machine translation. However, the results were poor in the late 1950s, and a working system was yet to be shown. All work was therefore theoretical work implemented on a computer, which proved to be sub-optimal.

\section{The formation of the Croatian group in Zagreb}
\label{sec:1}

In Yugoslavia, organized effort in machine translation started in 1959, but the first individual effort was made by Vladimir Matkovi\' c from the Institute for Telecommunications in Zagreb in 1957 in his PhD thesis on entropy in the Croatian language \cite{FiknaLaszlo11}. The main research group in machine translation was formed in 1958, at the Circle for Young Linguists in Zagreb, initiated by a young linguist Bulcsu Laszlo, who graduated in Russian language, Southern Slavic languages and English language and literature at the University of Zagreb in 1952.
The majority of the group members came from different departments of the Faculty of Humanities and Social Sciences of the University of Zagreb, with several individuals from other institutions. The members from the Faculty of Humanities and Social Sciences were: Svetozar Petrovi\' c (Department of Comparative Literature), Stjepan Babi\' c (Department of Serbo-Croatian Language and Literature), Krunoslav Pranji\' c (Department of Serbo-Croatian Language and Literature), \v Zeljko Bujas (Department of English Language and Literature), Malik Muli\' c (Department of Russian Language and Literature) and Bulcsu Laszlo (Department of Comparative Slavistics). The members of the research group from outside the Faculty of Humanities and Social Sciences were: Bo\v zidar Finka (Institute for Language of the Yugoslav Academy of Sciences and Arts), Vladimir Vrani\' c (Center for Numerical Research of the Yugoslav Academy of Sciences and Arts), Vladimir Matkovi\' c (Institute for Telecommunications), Vladimir Muljevi\' c (Institute for Regulatory and Signal Devices) \cite{FiknaLaszlo11}.

Laszlo and Petrovi\' c \cite{Uvod12} also commented on the state of the art of the time, noting the USA prototype efforts from 1954 and the publication of a collection of research papers in 1955 as well as the USSR efforts starting from 1955 and the UK prototype from 1956. They do not detail or cite the articles they mention. However, the fact that they referred to them in a text published in 1959 (probably prepared for publishing in 1958, based on \cite{Uvod12}, where Laszlo and Petrovi\' c described that the group had started its work in 1958) leads us to the conclusion that the poorly funded Croatian research was lagging only a couple of years behind the research of the superpowers (which invested heavily in this effort). Another interesting moment, which they delineated in \cite{Uvod12}, is that the group soon discovered that some experimental work had already been done in 1957 at the Institute of Telecommunications (today a part of the Faculty of Electrical Engineering and Computing at the University of Zagreb) by Vladimir Matkovi\' c. Because of this, they decided to include him in the research group of the Faculty of Humanities and Social Sciences at the University of Zagreb. The work done by Matkovi\' c was documented in his doctoral dissertation but remained unpublished until 1959.

\noindent The Russian machine translation pioneer Andreev expressed hope that the Yugoslav (Croatian) research group could create a prototype, but sadly, due to the lack of federal funding, this never happened \cite{FiknaLaszlo11}. Unlike their colleagues in the USA and the USSR, Laszlo’s group had to manage without an actual computer (which is painfully obvious in \cite{Spalatin13}), and the results remained mainly theoretical. Appealing probably to the political circles of the time, Laszlo and Petrovi\' c note that, although it sounds strange, research in computational linguistics is mainly a top-priority military effort in other countries \cite{Uvod12}. There is a quote from \cite{FiknaLaszlo11} which perhaps best delineates the optimism and energy that the researchers in Zagreb had:
\begin{quote}
"[...] The process of translation has to mechanicalized as soon as possible, and this is only possible if a competent, fast and inexhaustible machine which could inherit the translation task is created, even if just schematic. The machine needs to think for us. If machines help humans in physical tasks, why would they not help them in mental tasks with their mechanical memory and automated logic" (p. 118).
\end{quote}

\section{Contributions of the Croatian group}
\label{sec:2}
Laszlo and Petrovi\' c \cite{Uvod12} considered cybernetics (as described in \cite{Wiener14} by Wiener, who invented the term “cybernetics”) to be the best approach for machine translation in the long run. The question is whether Laszlo's idea of cybernetics would drive the research of the group towards artificial neural networks. Laszlo and his group do not go into neural network details (bear in mind that this is 1959, the time of Rosenblatt), but the following passage offers a strong suggestion about the idea they had (bearing in mind that Wiener relates McCulloch and Pitts' ideas in his book): "Cybernetics is the scientific discipline which studies analogies between machines and living organisms" (\cite{Uvod12}, p. 107). They fully commit to the idea two pages later (\cite{Uvod12}, p. 109): "An important analogy is the one between the functioning of the machine and that of the human nervous system". This could be taken to mean a simple computer brain analogy in the spirit of \cite{PMc15} and later \cite{Putnam16}, but Laszlo and Petrovi\' c specifically said that thinking of cybernetics as the "theory of electronic computers" (as they are made) is wrong \cite{Uvod12}, since the emphasis should be on modelling analogical processes. There is a very interesting quote from \cite{Uvod12}, where Laszlo and Petrovi\' c note that "today, there is a significant effort in the world to make fully automated machine translation possible; to achieve this, logicians and linguists are making efforts on ever more sophisticated problems". This seems to suggest that they were aware of the efforts of logicians (such as Bar Hillel, and to some degree Pitts, since Wiener specifically mentions logicians-turned-cyberneticists in his book \cite{Wiener14}), but still concluded that a cybernetic approach would probably be a better choice.

Laszlo and Petrovi\' c \cite{Uvod12} argued that, in order to trim the search space, the words would have to be coded so as to retain their information value but to rid the representations of needless redundancies. This was based on previous calculations of language entropy by Matkovi\' c, and Matkovi\' c's idea was simple: conduct a statistical analysis to determine the most frequent letters and assign them the shortest binary code. So A would get 101, while F would get 11010011 \cite{Uvod12}. Building on that, Laszlo suggested that, when making an efficient machine translation system, one has to take into account not just the letter frequencies but also the redundancies of some of the letters in a word \cite{Las17}. This suggests that the strategy would be as follows: first make a thesaurus, and pick a representative for each meaning, then stem or lemmatize the words, then remove the needless letters from words (i.e. letters that carry little information, such as vowels, but being careful not to equate two different words), and then encode the words in binary strings, using the letter frequencies. After that, the texts are ready for translation, but unfortunately, the translation method is never explicated. Nevertheless, it is hinted that it should be "cybernetic", which, along with what we have presented earlier, would most probably mean artificial neural networks. This is highlighted by the following passage (\cite{Uvod12}, p. 117):

\begin{quote}
"A man who spends 50 years in a lively and multifaceted mental activity hears a billion and a half words. For a machine to have an ability comparable to such an intellectual, not just in terms of speed but also in terms of quality, it has to have a memory and a language sense of the same capacity, and for that - which is paramount - it has to have in-built conduits for concept association and the ability to logically reason and verify, in a word, the ability to learn fast."
\end{quote}

\noindent Unfortunately, this idea of using machine learning was never fully developed, and the Croatian group followed the Soviet approach(es) closely. Pranji\' c \cite{Pranj17} analyses and extrapolates five basic ideas in the Soviet Machine Translation program, which were the basis for the Croatian approach:
\begin{enumerate}
    \item Separation of the dictionary from the MT algorithm
    \item Separation of the understanding and generation modules of the MT algorithms
    \item All words need to be lemmatized
    \item The word lemma should be the key of the dictionary, but other forms of the word must be placed as a list in the value next to the key
    \item Use context to determine the meaning of polysemous words.
\end{enumerate}

\noindent The dictionary that was mentioned before is, in fact, the intermediary language, and all the necessary knowledge should be placed in this dictionary, the keys should ideally be just abstract codes, and everything else would reside and be accessible as values next to the keys \cite{Spalatin13}. Petrovi\' c, when discussing the translation of poetry \cite{Petr19}, noted that ideally, machine translation should be from one language to another, without the use of an intermediate language of meanings.

Finka and Laszlo envisioned three main data preparation tasks that are needed before prototype development could commence \cite{FiknaLaszlo11}. The first task is to compile a dictionary of words sorted from the end of the word to the beginning. This would enable the development of what is now called stemming and lemmatization modules: a knowledge base with suffixes so they can be trimmed, but also a systematic way to find the base of the word (lemmatization) (p. 121). The second task would be to make a word frequency table. This would enable focusing on a few thousand most frequent words and dropping the rest. This is currently a good industrial practice for building efficient natural language processing systems, and in 1962, it was a computational necessity. The last task was to create a good thesaurus, but such a thesaurus where every data point has a "meaning" as the key, and words (synonyms) as values. The prototype would then operate on these meanings when they become substituted for words.

But what are those meanings? The algorithm to be used was a simple statistical alignment algorithm (in hopes of capturing semantics) described in \cite{Spalatin13} on a short Croatian sentence "\v covjek [noun-subject] pu\v si [verb-predicate] lulu [noun-objective]" (A man is smoking a pipe). The first step would be to parse and lemmatize. 
Nouns in Croatian have seven cases just in the singular, with different suffixes, for example:\\
\v COVJEK - Nominative singular\\
\v COVJEKA - Genitive singular\\
\v COVJEKU - Dative singular\\
\v COVJEKA - Accusative singular\\
\v COVJE\v CE - Vocative singular\\
\v COVJEKU - Locative singular\\
\v COVJEKOM - Instrumental singular\\
Although morphologically transparent, the lemma in the mentioned case would be “\v COVJEK-”; there is a voice change in the Vocative case, so for the purpose of translation, “\v COVJE-” would be the “lemma”. The other two lemmas are PU\v s- and LUL-.\\
The thesaurus would have multiple entries for each lemma, and they would be ordered by descending frequency (if the group actually made a prototype, they would have realized that this simple frequency count was not enough to avoid only the first meaning to be used). The dictionary entry for \v COVJE- (using modern JSON notation) is:\\
"\v COVJE-": {"mankind": {193.5: "LITTLENESS", 690.2: "AGENT"}, "man": {554.4: "REPRESENTATION", 372.1: "MANKIND", 372.3: "MANKIND" ...}, ...}\\
The meaning of the numbers used is never explained, but they would probably be used for cross-referencing word categories.\\
After all the lemmas comprising the sentence have been looked up in this dictionary, the next step is to keep only the inner values and discard the inner keys, thus collapsing the list, so that the example above would become:\\
"COVJE-": {193.5: "LITTLENESS", 690.2: "AGENT", 554.4: "REPRESENTATION", 372.1: "MANKIND", 372.3: "MANKIND" ...}\\
Next, the most frequently occurring meaning would be kept, but only if it grammatically fits the final sentence. One can extrapolate that it is tacitly assumed that the grammatical structure of the source language matches the target language, and to do this, a kind of categorical grammar similar to Lambek calculus \cite{Lamb20} would have to be used. It seems that the Croatian group was not aware of the paper by Lambek (but only of Bar-Hillel's papers), so they did not elaborate this part.

Finka \cite{Finka21} notes that Matkovi\' c, in his dissertation from 1957, considered the use of bigrams and trigrams to “help model the word context”. It is not clear whether Finka means character bigrams, which was computationally feasible at the time, or word bigrams, which was not feasible, but the suggestion of modelling the word context does point in this direction. Even though the beginnings of using character bigrams can be traced back to Claude Shannon \cite{Sha22}, using character-level bigrams in natural language processing was studied extensively only by Gilbert and Moore \cite{GM23}. It can be argued, that in a sense, Matkovi\' c predated these results, but his research and ideas were not known in the west, and he was not cited. The successful use of word bigrams in text classification had to wait until \cite{Lew24}. The long time it took to get from character to words was mainly due to computational limitations, but Matkovi\' c's ideas are not to be dismissed lightly on account of computational complexity, since the idea of using word bigrams was being explored by the Croatian group--perhaps the reason for considering such an idea was the lack of a computer and the underestimation of the memory requirements. The whole process described above is illustrated in Fig. 1.

\begin{center}
  \includegraphics[width=8cm]{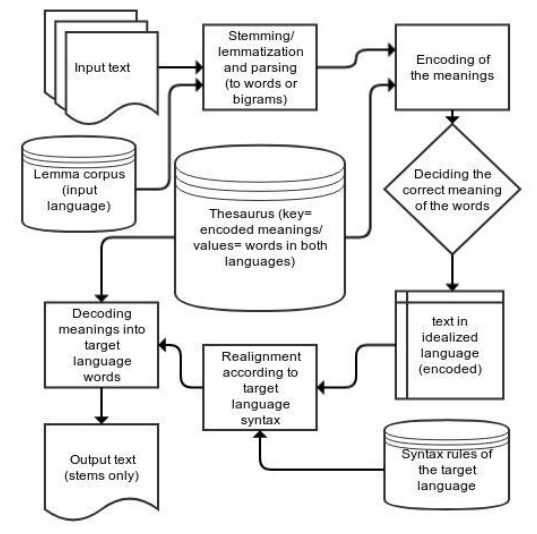}
%\caption{\textbf{Fig. 1.} Illustration of the translation process as envisioned by Laszlo}
\end{center}

\noindent Several remarks are in order. First, the group seemed to think that encodings would be needed, but it seems that entropy-based encodings and calculations added no real benefits (i.e. added no benefit that would not be offset by the cost of calculating the codes). In addition, Finka and Laszlo \cite{FiknaLaszlo11} seem to place great emphasis on lemmatization instead of stemming, which, if they had constructed a prototype, they would have noticed it to be very hard to tackle with the technology of the age. Nevertheless, the idea of proper lemmatization would probably be replaced with moderately precise hard-coded stemming, made with the help of the "inverse dictionary", which Finka and Laszlo proposed as one of the key tasks in their 1962 paper. This paper also highlights the need for a frequency count and taking only the most frequent words, which is an approach that later became widely used in the natural language processing community. Sentential alignment coupled with part-of-speech tagging was correctly identified as one of the key aspects of machine translation, but its complexity was severely underestimated by the group. One might argue that these two modules are actually everything that is needed for a successful machine translation system, which shows the complexity of the task.

As noted earlier, the group had no computer available to build a prototype, and subsequently, they have underestimated the complexity of determining sentential alignment. Sentential alignment seems rather trivial from a theoretical standpoint, but it could be argued that machine translation can be reduced to sentential alignment. This reduction vividly suggests the full complexity of sentential alignment. But the complexity of alignment was not evident at the time, and only several decades after the Croatian group's dissolution, in the late 1990s, did the group centered around Tillmann and Ney start to experiment with statistical models using (non-trivial) alignment modules, and producing state-of-the-art results (cf. \cite{Vog25}) and \cite{Och26}. However, this was statistical learning, and it would take another two decades for sentential alignment to be implemented in cybernetic models, by then known under a new name, deep learning. Alignment was implemented in deep neural networks by \cite{Beng27} and \cite{Ney28}, but a better approach, called attention, which is a trainable alignment module, was being developed in parallel, starting with the seminal paper on attention in computer vision by \cite{Mnih29}.

\section{Conclusion}
At this point, we are leaving the historical analysis behind to speculate on what the group might have discovered if they had had access to a computer. First of all, did the Croatian group have a concrete idea for tackling alignment? Not really. However, an approach can be read between the lines of primarily \cite{Las17} and \cite{Pranj17}. In \cite{Pranj17}, Prani\' c addresses the Soviet model by Andreev, looking at it as if it was composed of two modules – an understanding module and a generation module. Following the footsteps of Andreev, their interaction should be over an idealized language. Laszlo \cite{Las17} notes that such an idealized language should be encoded by keeping the entropy in mind. He literally calls for using entropy to eliminate redundancy while translating to an artificial language, and as Muli\' c notes \cite{Mulic8}, Andreev's idea (which should be followed) was to use an artificial language as an intermediary language, which has all the essential structures of all the languages one wishes to translate.

The step which was needed here was to eliminate the notion of structure alignment and just seek sentential alignment. This, in theory, can be done by using only entropy. A simple alignment could be made by using word entropies in both languages and aligning the words by decreasing entropy. This would work better for translating into a language with no articles. A better approach, which was not beyond the thinking of the group since it was already proposed by Matkovi\' c in his dissertation from 1957 \cite{Finka21}, would be to use word bigrams and align them.
It is worth mentioning that, although the idea of machine translation in the 1950s in Croatia did not have a significant influence on development of the field, it shows that Croatian linguists had contemporary views and necessary competencies for its development. But, unfortunately, the development of machine translation in Croatia had been stopped because of the previously discussed circumstances. In 1964, Laszlo went to the USA, where he spent the next seven years, and after returning to Croatia, he was active as a university professor, but because of disagreement with the ruling political option regarding Croatian language issues, he published very rarely and was mainly focused on other linguistic issues in that period, but his work was a major influence on the later development of computational linguistics in Croatia.

\bibliographystyle{plain}
\bibliography{Laszlo} 

\end{document}